\documentclass[12pt]{iopart}
\usepackage{graphicx}
\usepackage[T1]{fontenc}
\usepackage{lmodern}
\usepackage{amsfonts}

\begin{document}

\title{Genealogies in simple models of evolution}

\author{ \'Eric Brunet  and  Bernard Derrida}
\address{Laboratoire de Physique Statistique, 
\'Ecole Normale Sup\'erieure,
\\
UPMC, Universit\'e Paris Diderot, CNRS,
\\24, rue Lhomond, 75231 Paris Cedex 05,
France}
\ead{Eric.Brunet@lps.ens.fr}

\begin{abstract}
We review the statistical properties of the genealogies of a few models of evolution.
In the asexual case, selection leads to coalescence times which grow logarithmically with the size of the population in contrast with the linear growth of the neutral case.
Moreover  for a whole class of models, the statistics of the genealogies are those of the Bolthausen Sznitman coalescent rather than the Kingman coalescent in the neutral case.
For sexual reproduction, the time to reach the first common ancestors to the whole population and the time  for  all individuals to have all their ancestors in common are also logarithmic in the neutral case, as predicted by Chang \cite{Chang.99}. We discuss how these times are modified in a simple way of introducing selection. 
\end{abstract}

\section{Introduction  }
\label{intro}
The genealogy of a population describes  the relationships between  all the ancestors of this population.
Simple questions one may ask about the genealogy of a population are:
\begin{description}
\item How far one has to go into the past to find the most recent 
ancestor of two individuals? of $m$ individuals? of the whole population?
\item How do these times depend on the  sample   of $m$  individuals chosen at random  in the population?
\item How do they  depend on the size of the population?
\item How do they  fluctuate when the population evolves over many generations?
\item How are they affected by  the forces (like selection) acting on evolution?
\end{description}

\begin{figure}
\centering
\begin{indented}\item[]
\includegraphics[width=.75\textwidth]{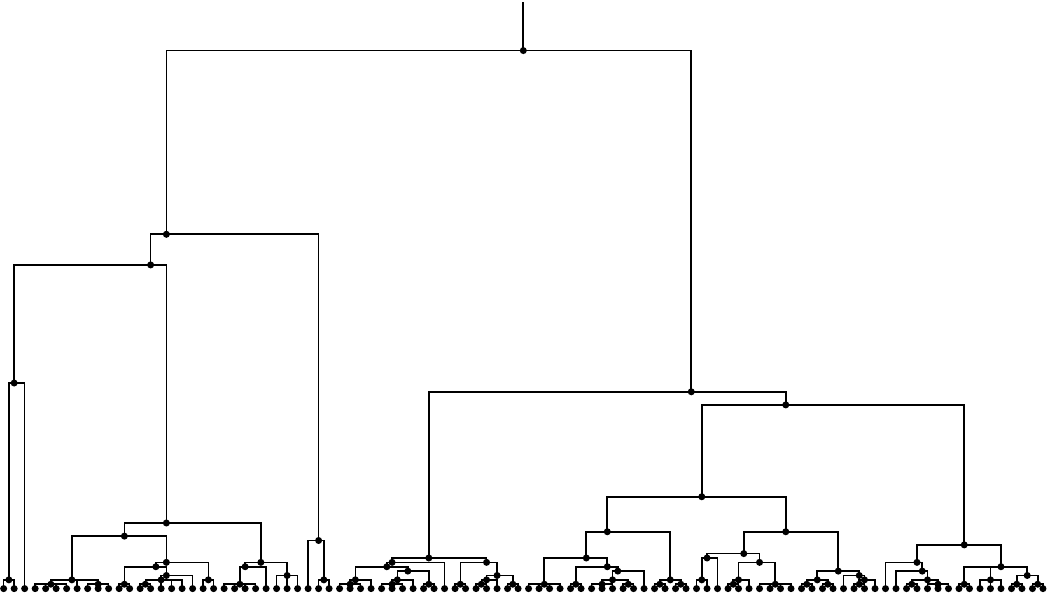}
\end{indented}
\caption{A typical genealogical tree in the neutral case, obtained by
simulating the Wright-Fisher model for a population of $N=100$ individuals. For the particular realization shown on  the figure, the number of generations to reach the most recent common ancestor is 125, which is, as expected,  of order $N$. Already for $N=100$, all the visible nodes are coalescences of pairs of branches and one cannot see any  multiple coalescences except at the very bottom of the figure where the number of branches is still of order $N$.}
\label{t1}
\end{figure}

In the case of {\it an  asexual evolution}, the ancestry of a population is a tree, the root of which is the most recent common ancestor of the whole population. 
In the neutral case { (i.e. when  all  individuals have on average the same number of surviving offspring at the next generation)}, for a well mixed population, 
the height of the tree is  proportional to the size of the population
(see figure \ref{t1}) and its   statistics  are  described by Kingman's
coalescent \cite{Kingman2.82,Kingman.82,Hudson.91,DonnellyTavare.95} (see section 2).

\begin{figure}
\centering
\begin{indented}\item[]
\includegraphics[width=.75\textwidth]{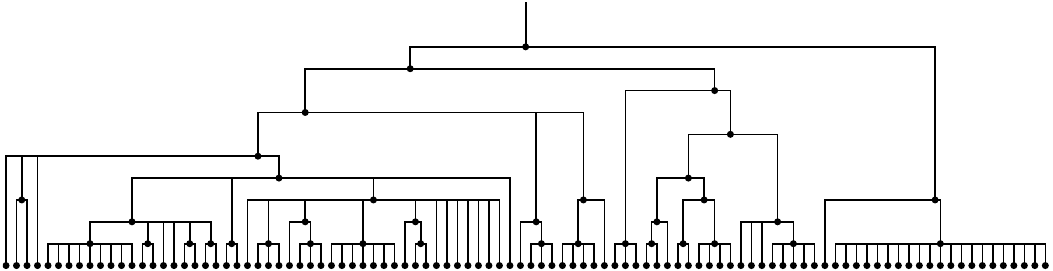}
\end{indented}
\caption{A typical genealogical tree in  presence of selection for a population of size $N=100$. The number  of  generations (10  for the realization of  the exponential model shown on  the figure)  to reach the most recent common ancestor is much shorter than in the neutral case.
In contrast to the neutral case, one can observe multiple coalescences even rather high in the tree.}
\label{t2}
\end{figure}

In a series of recent papers, together with A.H. Mueller and S. Munier
\cite{BDMM2.06,BDMM.07}, we considered a family of models of evolution
with selection. For these models, in contrast to the neutral case, the  height of the tree
grows logarithmically with the size of the population and,    its shape
(see figure \ref{t2}) is given asymptotically  by the Bolthausen-Sznitman
coalescent \cite{BolthausenSznitman.98} rather than by Kingman's
coalescent.

\begin{figure}
\begin{indented}\item[]
\includegraphics[width=.35\textwidth]{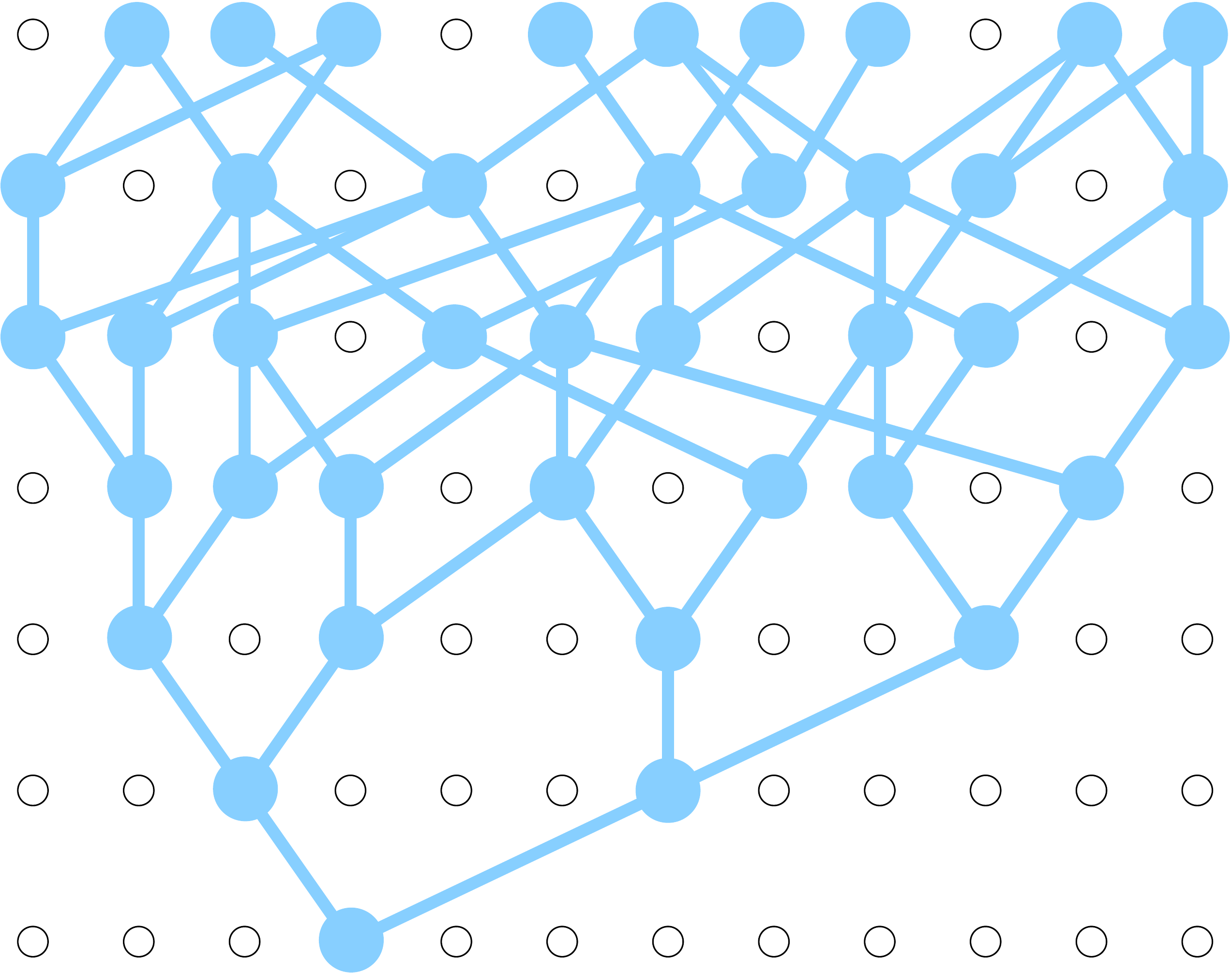} \qquad
\includegraphics[width=.35\textwidth]{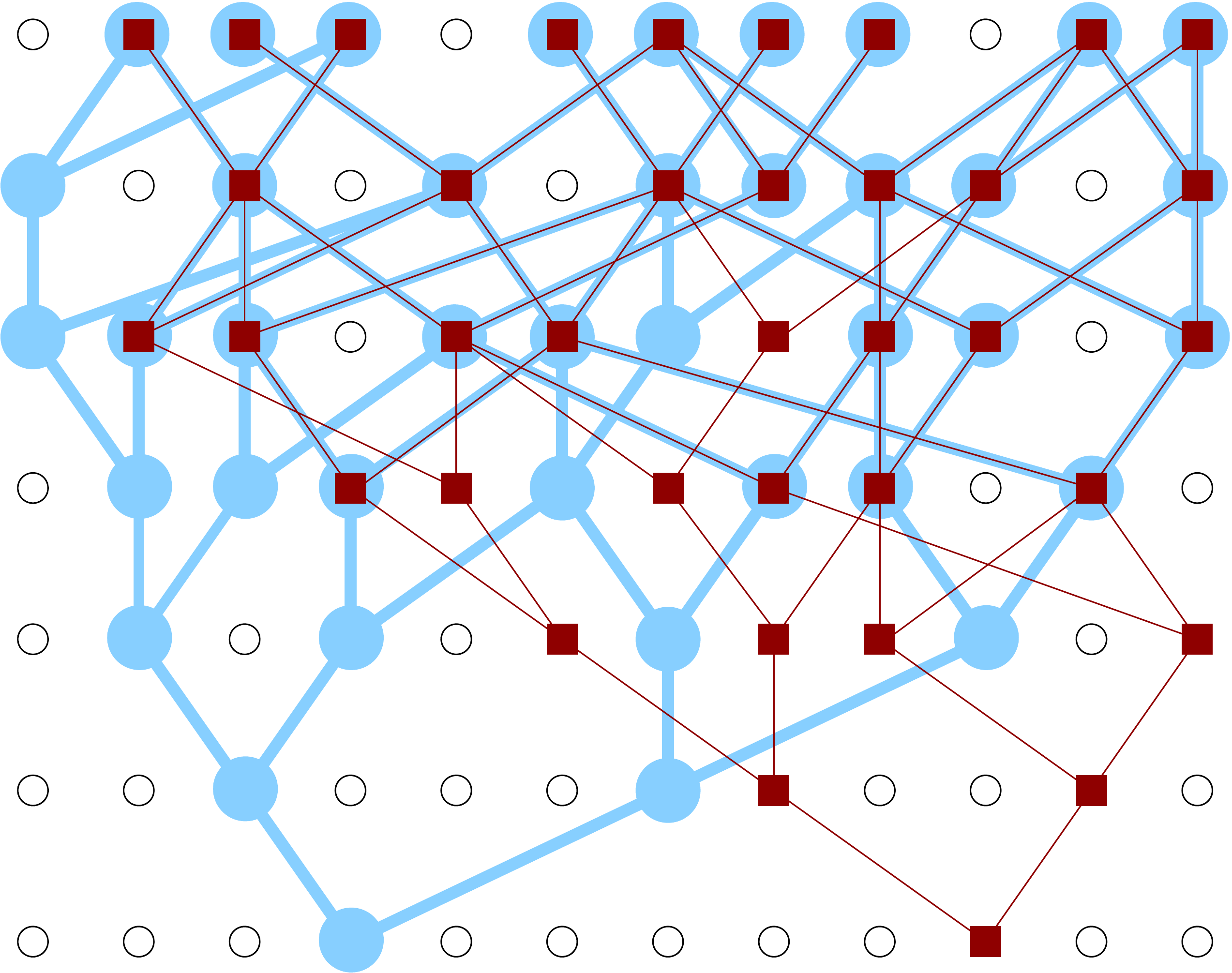}
\end{indented}
\caption{The ``tree'' of ancestors in the case of sexual reproduction.
After a few generations (of order the log of the size of the population),
the branches start to merge and the total number of ancestors saturates
(left figure). Furthermore, comparing the ancestries of two individuals (right figure),
all their ancestors become identical.}
\label{2p}
\end{figure}

In the case of {\it sexual evolution}, each individual has two parents, four grand parents, and so on.  Each individual is  therefore  the root of what looks like a tree for the first generations in the past. Going further into the past,  however, the branches of this tree start to merge (see figure \ref{2p})  and  the 
number of distinct ancestors  do not  grow exponentially anymore
\cite{Chang.99,DerridaManrubiaZanette.00}. The number of ancestors then saturates at  a value which is a fraction of the whole population living in this remote past (the 
rest of this past population consists  of all those individuals who had no
offspring or whose lineage became extinct). 
Comparing the ancestry of two individuals of the same generation, one sees two growing binary trees  in the recent past which then start to intermix in a more remote past until they become identical.

Here we try to review a few properties  of the genealogies of some simple models of  evolving   populations, both in the neutral case and in the case of selection.
In section~2, we recall a few properties of the Kingman coalescent and of the Bolthausen-Sznitman coalescent, as special cases of coalescent processes
and why neutral evolution in the asexual case leads to the Kingman coalescent, in the limit of a large population. In section~3, we describe  the properties of the exponential model,   an exactly soluble of asexual evolution in presence  of selection, and show that  its trees  follow the statistics of  the  Bolthausen-Sznitman coalescent.
In section~4, we show that, like the exponential model,  more generic
models of evolution  with selection converge  to the Bolthausen-Sznitman coalescent.
Section 5 is devoted to the case of sexual evolution both in the neutral case and in presence   of selection.

\section{How to quantify genealogies in the case of   asexual reproduction
}

To study the statistical properties of the  trees generated by sexual reproduction, one can adopt several points of view:

\begin{enumerate}
\item One can try to see how the population can be partitioned into $\tau$-families, with the rule that two individuals belong to the same $\tau$-family if and only if  their most recent common ancestor is at a distance less than $\tau$  generations in the past.
These $\tau$-families can themselves be divided into subfamilies by choosing a shorter  number of generations  $\tau'$, and so on. 

\item One can alternatively study the   genealogical tree as a dynamical process   and try to determine  the rates $q_k$ at which 
 $k$ branches merge into one when one moves from the bottom to the top of
the tree (\textit{i.e.}\@ when one looks at the tree backwards in time). These rates might be  correlated in time  or  depend on the  characteristics of the individuals along the branches which merge.

\item 
In our previous works \cite{BDMM2.06,BDMM.07,BrunetDerrida.12}
we chose to compute  for   these random trees     the coalescence times $T_p$ defined as follows:
$T_p$ is the age of the most recent common ancestor of $p$ individuals chosen at random in the population. In general the times $T_p$  depend both on  the sample of $p$ individuals chosen and on the generation. One advantage of characterizing the random trees  by the times $T_p$ is that they are  relatively easy to average    (over the samples of $p$ individuals and over the generations) in simulations.
\end{enumerate}
For some models, such as coalescence processes  discussed below,
one can find explicit mathematical formulae which relate these various properties \cite{BrunetDerrida.12}.
\subsection{Coalescence models}
\label{coal}
Coalescent processes give a simple procedure  to generate a whole class
of random trees
\cite{Nordborg.01,Berestycki.09,Pitman.99,DurrettSchweinsberg.05}. In a continuous time version, a coalescent process is a dynamical stochastic   process, where any set of $k$ individuals  have a probability  $q_k dt $  of coalescing into one individual during the infinitesimal time interval $dt$, implying that, 
if the number of branches is $b$  at time $t$, the probability  $r_{b}(b') dt$ of having  $b'$ branches at time $t+ dt$  (with  $b' < b$)
is given by \cite{BrunetDerrida.12}
\begin{equation}
r_{b}(b')  = \sum_{n=0}^{b'-1} {b! \over (b-b'+1)! \ (b'-1-n)! \ n!} (-1)^n q_{n+b-b'+1}.
\label{rb1}
\end{equation}

From the rates $q_k$, one can easily calculate the  times $\langle T_p \rangle$   (averaged over all the realizations of the coalescence process)
 by analyzing what happens in a steady state situation during an infinitesimal time interval $dt$ 
\begin{equation}
  \langle T_p \rangle =  dt + \langle T_p \rangle \left(1-  dt \sum_{b'< p} r_p( b')\right) + dt \sum_{2 \leq b' < p} r_p(b') \langle T_{b'} \rangle + O\left(dt^2\right) . 
\nonumber
\end{equation}
One then gets  
\begin{eqnarray}
\langle T_2 \rangle = \frac{1 }{ q_2},  \qquad\qquad
\frac{\langle T_3 \rangle }{ \langle T_2 \rangle} = \frac{4 q_2 - 3 q_3
}{ 3 q_2 - 2 q_3}, \nonumber \\[-1ex] 
\label{T234}
\\[-1ex]
\frac{\langle T_4 \rangle }{ \langle T_2 \rangle} = \frac{27 q_2^2 - 56
q_2 q_3 + 28 q_3^2 + 12 q_2 q_4 - 10 q_3 q_4 }{ (3 q_2 - 2 q_3) ( 6 q_2 -
8 q_3 + 3 q_4)} .
\nonumber
\end{eqnarray}
So $q_2$ determines the  time scale of the times $T_p$ and all the ratios $\langle T_p \rangle / \langle T_2 \rangle$ are given by  ratios of the coalescing rates $q_k$. 

\goodbreak

{\it Remark:} an alternative way of thinking of the coalescence processes
\cite{Berestycki.09,DurrettSchweinsberg.05,HuilletMohle.12} defined above
is to say that  during every infinitesimal time interval $dt$, there is
a probability
$\rho(f) df dt$ that a fraction $ f$  of all the branches   coalesces into a single branch, all the other branches remaining unchanged. Then for $k$  given individuals, the probability that their $k$ branches merge during $dt$  is given by
\begin{equation}
q_k  \ dt = \int_0^1 f^k \rho(f) df  \  dt,
\label{fk}
\end{equation}
while the probability $r_b(b')dt $ that $b$  distinct branches  becomes $b'$ branches during the time interval $dt$ is
\begin{equation}
r_b(b') dt  = {b! \over (b'-1)! \  (b-b'+1)!} \int_0^1  f^{b-b'+1} (1-f)^{b'-1}  \rho(f) df \  dt
\nonumber
\end{equation}
and  one recovers (\ref{rb1}).

\bigbreak

Two special  cases will be of interest in what follows: 
\begin{enumerate}
\item {\it The Kingman coalescent } \\
In the Kingman coalescent,   only $q_2 \neq 0$ and all the $ q_k=0$ for $k \geq 3$. Then  (\ref{T234}) becomes
\begin{equation}
\langle T_2 \rangle = \frac{1 }{ q_2},  \qquad\qquad
\frac{\langle T_3 \rangle }{ \langle T_2 \rangle} = \frac{4
}{ 3 },  \qquad\qquad
\frac{\langle T_4 \rangle }{ \langle T_2 \rangle} = \frac{3 }{ 2 } .  \label{kingman}
\end{equation}
In fact all the   correlation functions between these times $T_p$ can be
computed  \cite{Tavare.84,DonnellyTavare.95,Berestycki.09,SimonDerrida.06}. In particular one  gets $ {\langle T_p \rangle }/{ \langle T_2 \rangle} = 2 -2/  p $.
\item {\it The  Bolthausen-Sznitman coalescent} \\
The Bolthausen-Sznitman coalescent \cite{BolthausenSznitman.98} was introduced in the context of the  mean field theory of spin glasses 
\cite{SherringtonKirkpatrick.75,KirkpatrickSherrington.78} to represent the tree structure 
of the pure states \cite{MezardPSTV.84,MezardParisiVirasoro.87} predicted by the replica scheme   invented by Parisi \cite{Parisi.79,Parisi.80,Parisi.83}.
In terms of the rates $q_k$,   the Bolthausen-Sznitman coalescent can be defined by
\begin{equation}
q_k= {q_2 \over k-1}
\label{qk-BS}
\end{equation}
leading to 
\begin{equation}
\langle T_2 \rangle = \frac{1 }{ q_2},  \qquad\qquad
\frac{\langle T_3 \rangle }{ \langle T_2 \rangle} = \frac{5
}{ 4 },  \qquad\qquad
\frac{\langle T_4 \rangle }{ \langle T_2 \rangle} = \frac{25 }{ 18 } .  \label{bs}
\end{equation}
\end{enumerate}
\bigbreak

A  natural question, then, is  to know whether a given model of evolution (with or without  selection)  gives rise to random trees which can be 
 described by   a particular coalescent process
 \cite{Nordborg.01,DurrettSchweinsberg.05,BDMM.07,HuilletMohle.12,BrunetDerrida.12}.
\subsection{The Wright-Fisher model}
The Wright-Fisher model  \cite{Nordborg.01,
Griffiths.80,DurrettSchweinsberg.05,MorelTeissierEtheridge.11}  is
one of the simplest and most studied models of evolution in the neutral  case. 
In its simplest version, it describes a population of fixed size $N$. At every generation $g$, the  parent of each individual is chosen uniformly
among the $N$ individuals living at the previous generation. 
It is easy to see that the probability $q_k$ that $k$ individuals have the same parent at the previous generation is $q_k= N^{1-k}$, so that $q_k \ll q_2$ for $k>2$ and large $N$. Moreover  the probability of seeing more than one coalescence event among $k$ individuals (with $k $ of order $1$)  at a given generation becomes also much smaller than $q_2$ for large $N$. Therefore 
in the limit of a large population, the time   $\langle T_2 \rangle= q_2^{-1}$  scales like  $N$\begin{equation}
\langle T_2 \rangle \sim N
\end{equation}  the ratios $q_k /q_2 \to 0$ for $k \ge 3$, and  the statistics of the trees are given by the Kingman coalescent.

The Kingman coalescent is particularly central in the theory of
neutral evolution because it is universal
\cite{Griffiths.80,Tavare.84,DurrettSchweinsberg.05,DonnellyTavare.95,Nordborg.01}: one can change the rules in
the definition of the Wright-Fisher model in many ways   and in
the limit of a large population, one always recovers the Kingman coalescent.
For example one may choose the  parent~$i$  in the previous generation  with  a  non uniform probability $p_i$ and as long as these $p_i$ decay fast enough with the size $N$    of the population, one has in the large $N$ limit  $q_k / q_2 = (\sum_i p_i^k)/ (\sum_i p_i^2) \to 0$  for $k \ge 3$ and one recovers the Kingman coalescent
\cite{MohleSagitov.01}.
\\ \ \\
{\it Remark 1:}
Starting from the random tree structure of the Kingman coalescent,  and assuming mutations arising at constant rate  along the branches of these trees, one can predict the statistical properties of the genetic diversity
\cite{Ewens.72,Tajima.83,FuLi.97,DegnanSalter.05,DerridaPeliti.91,RosenbergNordborg.02,PrignanoServa.09}. 
\\ \ \\
{\it Remark 2:}
Even in the large $N$ limit,  the shape of the tree and the times $T_p$
fluctuate  when one  follows  the same population over many generations.
The correlations between these times at different  generations can be
computed \cite{Serva.05,SimonDerrida.06}. These fluctuating shapes of the
trees are very reminiscent of the tree structure of pure states predicted
by the mean field theory of spin glasses
\cite{MezardPSTV.84,MezardParisiVirasoro.87,DerridaPeliti.91,Derrida.97}.
\\ \ \\
{\it Remark 3:}
One can define a finite dimensional version  of the Wright-Fisher model by considering that the individuals of the population are at the nodes
of a  lattice of $L^d$ sites with periodic boundary conditions in dimension~$d$   and that each individual has its parent chosen at random among all the sites at a distance less than $l$ 
with $l \ll L$. 
\cite{Cox.89,LimicSturm.06}.
In dimension $ d \ge 2$, the genealogies of such populations have their
statistics  still given by the Kingman coalesent (the problem  can be
formulated as a reaction diffusion probleme whose upper critical
dimension is $2$).  On the other hand, in dimension~$1$, the statistical
properties of the trees are modified (they can be understood in terms of
coalescing random walks) and  (\ref{kingman}) is then replaced by
\cite{BrunetDerridaSimon.08} 
\begin{equation}
\langle T_2 \rangle \sim N^2,   \qquad\qquad
\frac{\langle T_3 \rangle }{ \langle T_2 \rangle} = \frac{7
}{ 5 },  \qquad\qquad
\frac{\langle T_4 \rangle }{ \langle T_2 \rangle} = \frac{8 }{ 5 } . 
\label{1d}
\end{equation}
Note that  in this case  the spatial aspect is crucial,  and   there is no choice of the $q_k$ for which the one dimensional problem  could  be reduced to a coalescent 
process as defined in section (\ref{coal}): for all the coalescent
processes, the age $T_2$ of the most recent common ancestor of two
individuals has an exponential distribution implying that $\langle T_2^2
\rangle / \langle T_2 \rangle^2=2 $ whereas
  \cite{BrunetDerridaSimon.08}, for the finite dimensional
model in $d=1$, this ratio takes the   value $12/5$.

\section{The exponential model: an exactly soluble model of evolution  with selection}
\label{expo-mod}
\subsection{Definition}
The exponential model \cite{BDMM2.06,BDMM.07,BrunetDerrida.12} is a
simple generalization  of the Wright-Fisher model, which includes the
effect of selection.    As in the Wright-Fisher model the size $N$  of
the  population remains the same at every generation. All the individuals
at a given generation are however not equivalent: each individual $i$ at
generation $g$  carries a value $ x_i(g)$ which represents a trait (or a
fitness in the sense used by Bak and Sneppen \cite{BakSneppen.93}).
There are then two steps
to go from generation $g$ to generation $g+1$:

{\it The reproduction step:}  each individual has  its offspring generated by a Poisson process of density $e^{-(x-x_i(g))}$. This simply means that, with   probability   $e^{-(y-x_i(g))} dy$,  there is an offspring of $x_i(g)$ in the infinitesimal interval   $(y,y+dy)$. 
 Repeating the procedure for all individuals $i$ at generation $g$, one gets that way  an infinite number of offspring along the real axis, with many  offspring for large negative $y$ and no offspring for sufficiently large positive $y$ (note that at the right of any  point $y$ on the real axis there are always a finite number of offspring).

 {\it The selection step:} at generation $g+1$, one keeps only the $N$ rightmost points   among all these offspring.
\\ \ \\ \ 

The simplicity of the exponential model  comes from the following identity 
\begin{equation}
\sum_{1 \leq i \leq N} e^{-(x-x_i(g) ) } = e^{-(x-X_{g})} \ \ \ \  \mbox{with } \ \ \ \ \
X_g =\ln \left[ \sum_{1 \leq i \leq N} e^{x_i(g)}  \right].
\label{Xg}
\end{equation}
This means that  the offspring of all the individuals at generation $g$ can be generated by a single Poisson process centered at position $X_g$. So  $X_g$ is the only information about generation $g$ needed to generate the next generation. 
 A simple consequence is that the successive shifts 
$X_{g+1} - X_g$
 of the position of the population are i.i.d.\@ variables. 
Another advantage of the exponential model  is that the population at generation $g$ consists of  the $N$ rightmost points of  a single Poisson process (\ref{Xg}) centered at $X_g$.  One can then show \cite{BrunetDerrida.12} that the population at generation $g+1$ can be generated in the following way:
 \begin{equation}
x_i(g+1) =  X_g+ Y_{g+1} + y_i(g+1),
 \end{equation}
where $Y_{g+1}$ and the  $N$ variables $y_i(g+1)$ are independent random variables distributed according to
\begin{eqnarray}
{\rm Prob }(Y < Y_{g+1}  < Y+ d Y ) = {1 \over N!}  \exp\left[- (N+1) Y -
e^{-Y }  \right] d Y ,
\nonumber
\\
{\rm Pro b}(y < y_i(g+1)  < y+ d y ) = \left\{  \begin{array}{ll} e^{-y }
\  d y  & {\rm for } \  y >0 ,
\\ 0 & {\rm for}  \ y <0 .\end{array} \right.
\label{Pro}
 \end{eqnarray}
 The positions $\{ x_i(g)\}$ form a cloud of points which remain  grouped as $g$ increases.  One could locate this cloud of points by its center of mass $X_g^{\rm C.M.}$.  As the points remain grouped, one can as well use the position $X_g$ defined in (\ref{Xg}) which has the advantage that the differences  
\begin{equation} 
X_{g+1} - X_g = Y_{g+1} + \ln \left[  \sum_i e^{y_i(g+1)}\right]
\end{equation} 
are uncorrelated random variables. It is easy to see that  the difference
\begin{equation}
X_g-X_g^{\rm C.M.} = \ln  \left[\sum_i e^{y_i(g)} \right] -  {1 \over N} \sum_i y_i(g)
\nonumber
\end{equation}
has  a distribution independent of $g$, so that all the cumulants of  $X_g^{\rm C.M.}$ and of  $X_g$ have the same linear growth with $g$.

For example the speed of adaptation $v_N$ (which is simply  the velocity of the cloud of points along the real axis)  is given by
 \begin{equation} 
\fl v_N= \lim_{g \to \infty} {\langle X_g \rangle \over g }
= \lim_{g \to \infty} {\langle X_g^{\rm C.M.} \rangle \over g } =  \langle
X_{g} - X_{ g-1} \rangle= \langle Y_g \rangle + \left\langle \ln
\left[\sum_i e^{y_i(g)} \right]  \right\rangle.
\label{velocity}
 \end{equation} 
Similarly 
the generating function $G_N(\beta)$ of all the cumulants of the positions $X_g$ or   $X_g^{\rm C.M.}$ can  be computed by
\begin{eqnarray}
G_N(\beta) & =   \lim_{g \to \infty} {\ln \langle e^{-\beta X_g }\rangle \over g }
= \lim_{g \to \infty} {\ln  \langle e^{-\beta X_g^{\rm C.M.} }\rangle
\over g } ,
\nonumber
\\ & =  \ln   \langle e^{-\beta  (X_g-X_{g-1}) }\rangle= \ln  \langle e^{-\beta Y_g
}\rangle +  \ln \left\langle \left[\sum_i e^{y_i(g)} \right]^{-\beta}
\right\rangle.
\label{generating}
 \end{eqnarray}

The average over $Y_g$ and the $y_i(g)$'s (distributed according to (\ref{Pro}))
can be performed leading to the following  
 large $N$ behavior   \cite{BDMM2.06,BDMM.07} 
\begin{equation}
 \fl G_N(\beta) \simeq -\beta  \ln  \ln  N - {1 \over \ln  N} \left( \beta \ln
\ln  N + \beta  \Gamma'(1) + \beta  -  \beta{\Gamma'(1+\beta) \over
\Gamma(1+\beta)} \right) + \cdots \  .
\label{Gbeta3}
\end{equation}
This gives for large $N$
\begin{eqnarray}
v_N = \lim_{g \to \infty} {\langle X_g \rangle \over g }
= \ln  \ln  N + {\ln  \ln  N  \ + \ 1 \over \ln  N} + \cdots\\
\lim_{g \to \infty} {\langle X_g^k \rangle_c \over g } =
  {  k!  \over \ln  N} \left[ \sum_{i \ge 1} {1 \over i^{k} }\right] + \cdots
\end{eqnarray}

This logarithmic growth of the velocity with the size of the population
seems to be  common to many models of evolution with selection on a
smooth landscape
\cite{Kessler.97,DesaiFisherMurray.07,RouzineBrunetWilke.08}  where each indivdual has a number of offspring proportional to the exponential of its fitness.

\subsection{The genealogies in the exponential model}
In the exponential model, given that  at generation $g+1$ there is an individual at position $x$,  the probability $p_i(x,g+1)$ that this  individual   is the offspring of the $i$-th individual  at generation $g$ is given by
\begin{equation}
p_i(x,g+1) = {e^{ -(x-x_i(g))} \over \sum_j e^{ -(x-x_j(g))} }
 = {e^{ x_i(g)} \over \sum_j e^{ x_j(g)} } =  {e^{ y_i(g)} \over \sum_{j
 } e^{ y_j(g)} }.
\end{equation}
The simplicity of the exponential model comes  from the fact that   these $p_i$ do not depend on $x$ and that they can be expressed in terms of the $y_i(g)$ only, implying that  the $p_i$ at different generations are uncorrelated.

The probability  $q_k$ that $k$ individuals have a single common ancestor at the previous generation  is then given by
\begin{equation}
q_k = \left\langle  \sum_i p_i^k\right\rangle    = \left\langle { \sum_{i
} e^{  k y_i(g)} \over \left[\sum_{j } e^{ y_j(g)} \right]^k}
\right\rangle  ,
\end{equation}
where the average is over the $N$ random variables $y_i(g)$ distributed according to (\ref{Pro}).
For large $N$ these averages can be computed \cite{BrunetDerrida.12}  
\begin{equation}
q_k \simeq { 1 \over (k-1) \ln  N}  . \label{qk-expo}
\end{equation}
Moreover, at each generation,    the probability of    more than one coalescence event
among a fixed number of individuals  becomes negligible in the large $N$ limit \cite{BrunetDerrida.12}.
Therefore the exponential model  converges to the Bolthausen-Sznitman coalescent (\ref{qk-BS}) with coalescence times given by  (\ref{bs})
and 
\begin{equation}
\langle T_2 \rangle \sim
 \ln  N.
\label{T2-expo}
\end{equation}
\subsection{Conditionning on the speed}
The shapes and the heights  of the trees fluctuate with $g$. In \cite{BrunetDerrida.12} we tried to determine how these fluctuations are correlated to the displacement
$X_g - X_0$. To do so we   assigned to each evolution event a weight $e^{-\beta X_g}$.
This means that, if $\overline{T_k(g)}$ is the age of the most recent ancestor  averaged over all the choices of $k$ individuals  at generation $g$ (for one realization of the process), we tried to compute  weighted averages such as
\begin{equation}
\langle T_k \rangle_\beta= \lim_{g\to \infty} {1 \over g}  \sum_{g'=1}^g
{\langle e^{-\beta X_g}  \overline{T_k(g')}  \rangle \over \langle
e^{-\beta X_g}  \rangle}.
\end{equation}
One can then show  \cite{BrunetDerrida.12} that, with these weights, the
probability that $k$ individuals at generation $g+1$ have their most recent common ancestor at the previous generation is 
\begin{equation}
q_k = {\left\langle  \sum_{i } e^{  k y_i(g)} \left[\sum_{j } e^{ y_j(g)}
\right]^{-\beta-k} \right\rangle \over \left\langle   \left[\sum_{j } e^{
y_j(g)} \right]^{-\beta} \right\rangle  }.
\label{qk-beta}
\end{equation}

As for $\beta=0$,  for large $N$, the probability of observing more than one coalescence event at each generation becomes negligible
and  (\ref{qk-beta}) gives to leading order \cite{BrunetDerrida.12} 
\begin{equation}
q_k= {1 \over \ln  N}{(k-2)! \  \Gamma(\beta+1) \over \Gamma(\beta+ k) }.
\label{qbeta}
\end{equation}
Therefore for these biased events, the trees have the same  statistics as a coalescent whose rates are given by (\ref{qbeta}).
This allows one to determine through (\ref{T234}) the coalescence times  $\langle T_p \rangle$ (see \cite{BrunetDerrida.12}  for explicit expressions). 

Varying $\beta$ in  (\ref{qbeta}) we see that,  conditioning on the displacement $X_g$, the statistics of the trees  interpolate between the Bolthausen-Sznitman coalescent for $\beta=0$  (no bias) and  the Kingman coalescent for $\beta \to \infty$ (bias toward slow velocities).

\section{ More generic  models of evolution with selection}
In \cite{BDMM2.06,BDMM.07,BrunetDerrida.12}, we considered  a  whole family of models of evolution with selection which generalize the exponential model. These more generic models are defined as follows: as in the exponential model, the population has a fixed size $N$   and each individual $i$  at generation $g$ is characterized by a real number  (a trait) 
 $x_i(g)$.  
At the next generation, the trait $x_i(g+1)$  of an offspring $i$  is inherited   from its parent $P(i,g+1)$ up to  a random number $z_{i}(g+1)$ drawn from a fixed  distribution $\rho(z)$ which represents the effect of the mutations on the trait $x_i$
\begin{equation}
x_i(g+1)= x_{P(i,g+1)}(g) + z_i(g+1) . 
\end{equation}
 Then comes the selection step where only  $N$ survivors among  all the offspring  produced by generation $g+1$ are chosen according to their trait $x_i(g+1)$.

The model  \cite{BDMM2.06,BDMM.07} depends on the number of offspring of each individual, on the distribution $\rho(z)$, and on the way the $N$ survivors at generation $g+1$ are chosen among all the offspring of generation $g$.

Here we present  the result of numerical simulations of three versions of the model, where we always took for $\rho(z)$ a flat distribution:

\begin{itemize}

\item {\it The perfect selection case:} in this version (see figure \ref{perfect-selection}), each individual  at generation $g$ has two offspring, but out of the
$2N$ resulting individuals we only keep at generation $g+1$  the $N$ rightmost ones  \textit{i.e.}\@ those who have
the highest $x_i(g+1)$. 
\begin{figure}
\begin{indented}\item[]
\includegraphics[width=.6\textwidth]{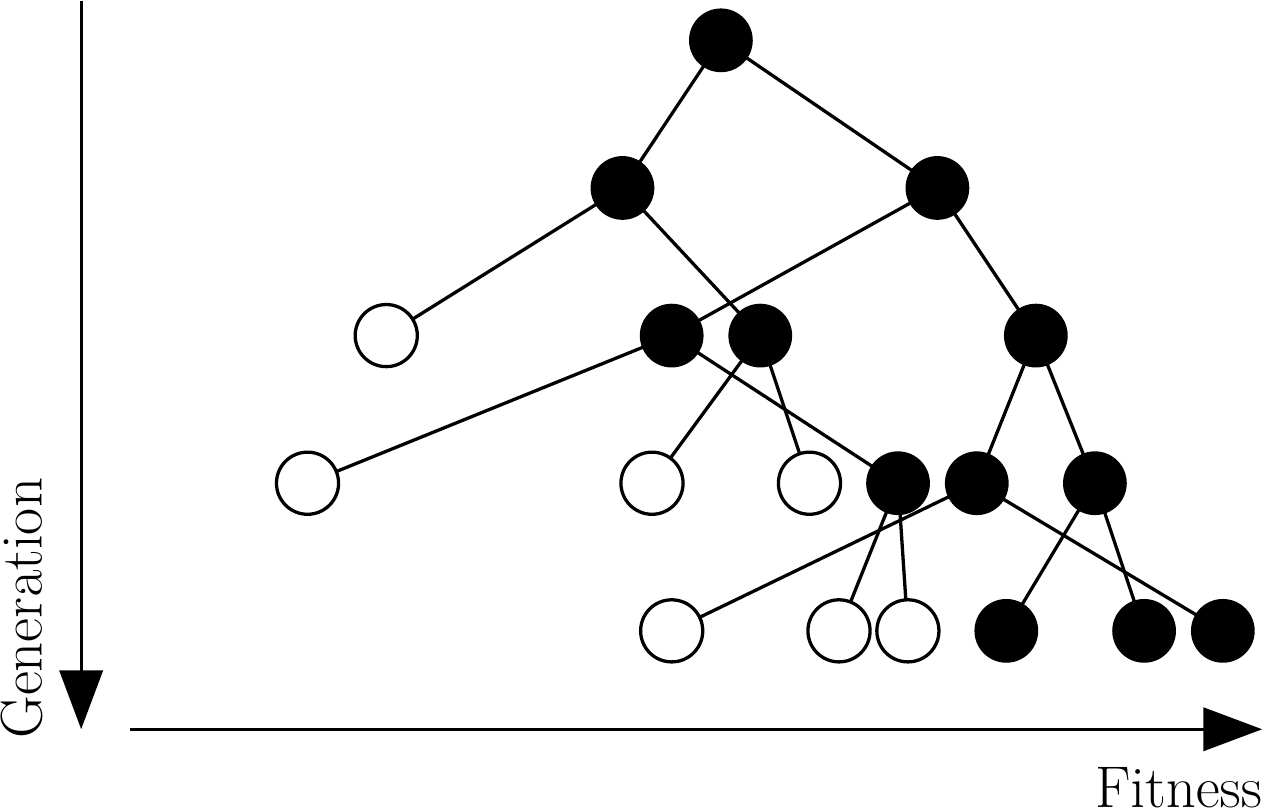}
\end{indented}
\caption{The perfect selection case:   at each generation one selects the $N$ (here $N=3$) rightmost offspring produced by the previous generation.
}
\label{perfect-selection}
\end{figure}

\item {\it The fuzzy  selection case:}  here again,  each individual has two offspring  at the next generation, but out of the
$2N$ resulting individuals,  the  $N$  survivors  at generation $g+1$ are chosen uniformly among
the $3N/2$ rightmost ones. 
\item{\it  The two parent selection case:}  in this version, for  each individual  at generation $g+1$, instead of choosing its parent
uniformly in the population  at generation $g$ as in the Wright-Fisher model, we choose 
 \emph{two} potential parents uniformly  in the population and keep as the
actual parent the best one among these two
\cite{YuEtheridge.08}. (It is equivalent to state that the $i$-th best individual has on
average $2(N-i)/(N-1)$ offspring and then draw the new generation with a
multinomial  distribution.)
\end{itemize}

\begin{figure}[ht]
\begin{indented}\item[]
\includegraphics[width=.75\textwidth]{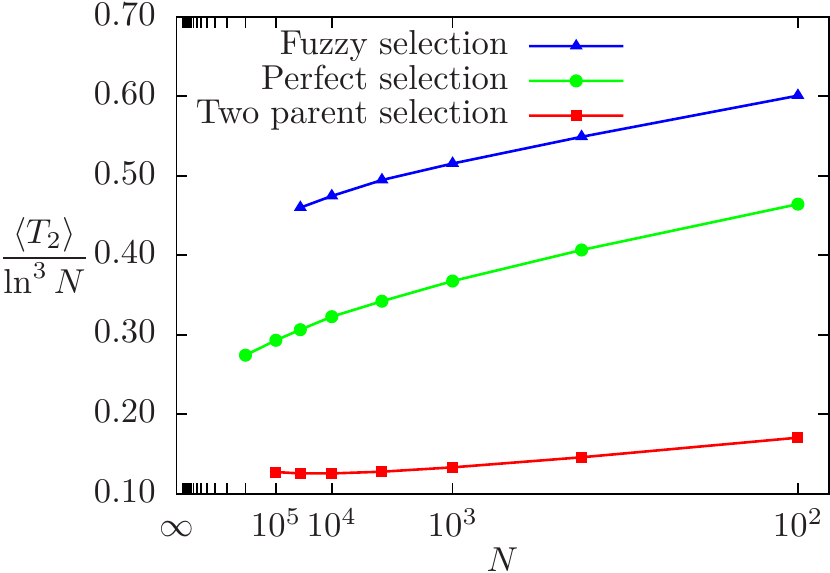}
\end{indented}
\caption{$\langle T_2\rangle/\ln^3N$ as a function of $N$ for the three
models. The scale on the horizontal
axis is $1/\ln^2N$.}
\label{figBST2}
\end{figure}

\begin{figure}[ht]
\begin{indented}\item[]
\includegraphics[width=.75\textwidth]{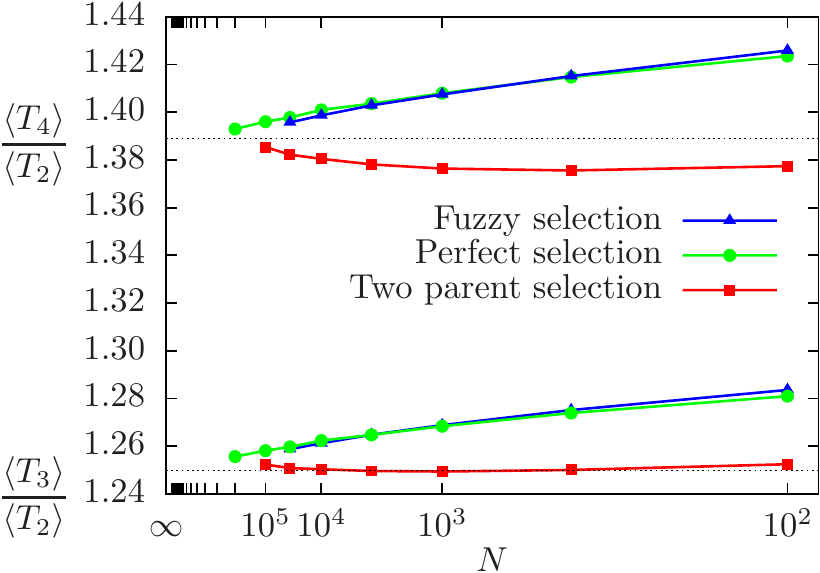}
\end{indented}
\caption{$\langle T_4\rangle/\langle T_2\rangle$ (up) and $\langle
T_3\rangle/\langle T_2\rangle$ (down) as a function of $N$ for the three
models. The dotted lines are the values for the Bolthausen-Snitzman
coalescent, respectively $25/18$ and $5/4$. The scale on the horizontal
axis is $1/\ln^2N$.}
\label{figBSratio}
\end{figure}

We measured the average coalescence times $\langle T_2\rangle$, $\langle
T_3\rangle$ and $\langle T_4\rangle$. The time $\langle T_2\rangle$
seems to increase like $\ln^3N$ (see figure~\ref{figBST2}) and the ratios
of these times seem to converge to the values
predicted for the Bolthausen-Snitzman coalescent in the large~$N$ limit,
(see figure~\ref{figBSratio}).

The above numerical simulations and additional ones (some on larger scales)
\cite{BDMM2.06,BDMM.07,BrunetDerrida.12,BDMM.06}
indicate that   for this whole family of models
the  statistics of  the genealogical trees  converge, in the large $N$ limit,   to those of the Bolthausen-Sznitman coalescent  (\ref{bs}) as for the
the exponential model, the only difference being that
the coalescence times $\langle T_2 \rangle$
grow like 
\begin{equation}
\langle T_2 \rangle \sim \ln^3 N.
\label{ln3}
\end{equation}
instead of  $\ln N$  for  the exponential model  (\ref{T2-expo}).
 
At each generation, the $N$ values $x_i(g)$ form a cloud of points which
moves  along the real axis as $g$ increases. This motion is stochastic and
we argued in \cite{BDMM2.06,BDMM.07}  that  its evolution can be related,
for large $N$, to that of a noisy Fisher-KPP equation in the weak noise
limit. In our joint works with A.H. Mueller and S. Munier
\cite{BDMM2.06,BDMM.07} we used a phenomenological   theory  \cite{BDMM.06}
developed for  such travelling wave equations   to explain the convergence
to the Bolthausen-Snitzman coalescent and the   $\ln^3N$ timescale. From
this phenomenological theory the following picture emerges: the evolution
of the cloud of points (the  $x_i(g)$) is most of the time  deterministic.
At time intervals of order $\ln^3 N$,  rare fluctuations  occur where  the
best individual gets an exceptionnally good trait $x_i(g)$. These
fluctuations relax on a time scale of order $\ln^2 N$ generations. During
this relaxation time, the individuals  with the  exceptionnally good trait
have their long time lineage less affected by selection than the rest of
the population and the net effect is  that their lineage replaces
a positive fraction~$f$ of the whole population as in (\ref{fk}); the value
of $f$ depends on the size of the fluctuation. An analysis of the
distribution of these rare fluctuations leads to a   distribution of $f$,
in (\ref{fk}), which corresponds to the  Bolthausen-Snitzman coalescent.
So the picture resembles that of the model considered by Durrett and
Schweinsberg \cite{DurrettSchweinsberg.05} where the effect of selection
ends up giving rise to a coalescent with multiple collisions.

\begin{figure}[ht]
\begin{indented}\item[]
\includegraphics[width=.75\textwidth]{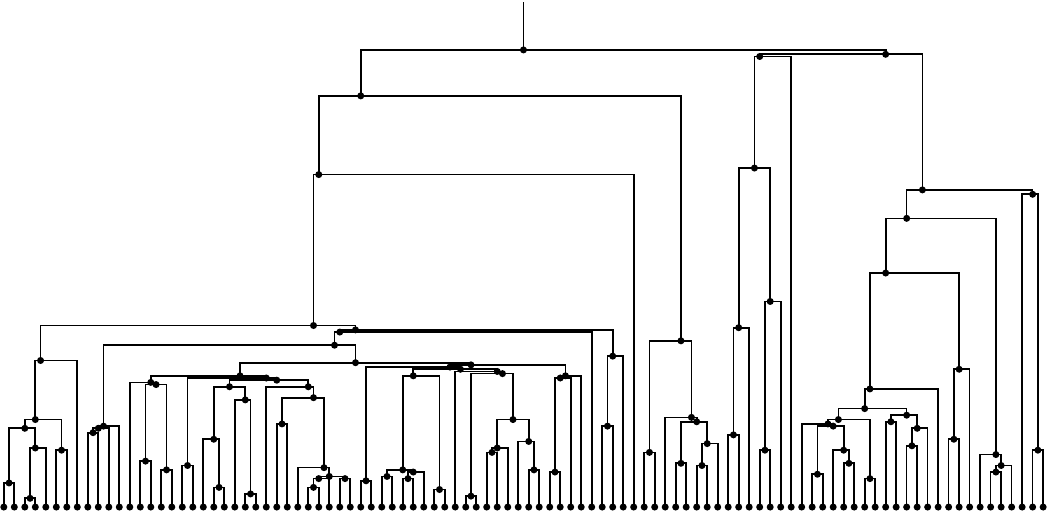}
\end{indented}
\caption{A realization of the genealogical tree of 100 individuals chosen
at random among a population of $10^4$ in the model with ``perfect
selection''.}
\label{fig100outof10000}
\end{figure}

It might look paradoxical   to get the Bolthausen-Sznitman coalescent, with multiple branches coalescing at the same point,  in situations like the perfect selection case, where each individual has only two offspring. This is simply because  coalescences of pairs  at  nearby generations look like multiple coalescences on a time scale of order $\ln^3 N$.
Figure~\ref{fig100outof10000} illustrates this  mechanism.
 We simulated the model with perfect selection  for a population of size  $N=10^4$
 and, at an arbitrary time, we plotted the coalescing tree of
the ancestries of $100$ individuals chosen at random. While 
a close look makes it apparent that there are only coalescences of pairs,
one can see several regions where coalescences
of pairs occur in quick succession, thus resulting into the
quasi-instantaneous coalescence of more than two lines
\cite{DurrettSchweinsberg.05}.
Note also that during the $\ln^3N$ generations (in real time) needed for
a fluctuation to occur, the ancestral lines of the individuals diffuse
within the whole population. As the size of this  cloud is of  order $\ln
N$ along the real axis, it takes $\ln^2N$ time steps to explore the whole
population and by the time the next fluctuation occurs, these lines have
swept many times over the whole population. This explains the perharps
counter-intuitive observation that the average time of coalescence between
several individuals is independent of the positions of these individuals.

\section{Sexual reproduction}

In the case of sexual reproduction, each individual has two parents in the
previous generation. When the population is large, its number $n_g$ of
distinct ancestors, $g$ generations in the past, starts to grow like $2^g$.
For any finite population, however, this number $n_g$  of distinct
ancestors has to saturate and the branches of the genealogical tree have to
merge if one goes far enough into the past (see figure \ref{2p}). Looking
at a given model of evolution, with sexual reproduction,  for a population
of fixed size, one may wonder :
\begin{enumerate}
\item How does $n_g$  depend on $g$? 
\item When  does  the saturation  occur?
\item  How far one has to go into the past to  find {\it some } common ancestors to    a group of  $p$ individuals?
\item  When does a group of  $p$ individuals have {\it  all} their ancestors in common?
\end{enumerate}
\subsection{The neutral case}
In the neutral case, all these questions have simple answers. For a simple
model discussed below, for example, Chang
\cite{Chang.99,RohdeOlsonChang.04} has shown that the number generations
${G_N}$  for all individuals to  have at least one common ancestor is  (for
large $N$) 
\begin{equation}
G_N \sim {\ln  N \over \ln 2} \sim 1.44 \ln  N ,
\label{G1}
\end{equation}
while the number generations  $\widetilde{G_N}$ needed for the whole population  to have {\it all} their ancestors in common is 
\begin{equation}
\widetilde{G_N} \sim \left( {1 \over \ln 2}  -{1 \over \ln  (2(1-x^* )) }
\right) \ \ln N \sim 2.55 \ln  N ,
\label{G2}
\end{equation}
where  $x^*\simeq0.8$ is the non zero solution of $x^*= 1 - \exp(-2 x^*)$.

The simplest model of neutral evolution to describe the genealogies in the case of sexual reproduction  is to take a population of fixed size $N$ and to assign to each individual two parents chosen at random in the previous generation (for simplicity we won't  make any distinction between males and females;
moreover, the two parents  of each individual being chosen independently in the previous generation, the model allows  these two parents  to coincide a
with probability $1/N$. All these simplifications are in fact unimportant as they do not affect the large $N$ behavior (\ref{G1},\ref{G2})).
In such a model, the number $n_g$ of distinct ancestors of a given individual, $g$ generations ago in the past, is a Markov process. The distribution of $n_{g+1}$, given $n_g$, can be written but it leads to a rather complicated formula which is in fact not that useful to understand the large $N$ behavior of the model.
Instead one can calculate the first moments of $n_{g+1}$
 by writing that
\begin{equation}
n_{g+1} = N - \sum_{i=1}^N y_i,
\end{equation}
where $y_i =0$ if individual $i$  at generation $g+1$ (in the past) is the ancestor of at least one of the $n_g$ ancestors at generation $g$ and $y_i=1$ otherwise
(i.e.  $y_i= 1$ if  $i$ has no offspring among the $n_g$ ancestors at generation $g$). For $i,j,k$ distinct, 
\begin{equation}
\fl
\langle y_i \rangle = \left(1 - {1 \over N} \right)^{2 n_g},\quad
\langle y_i  y_j \rangle = \left(1 - {2 \over N} \right)^{2 n_g},\quad
\langle y_i  y_j y_k\rangle = \left(1 - {3 \over N} \right)^{2 n_g},
\end{equation}
etc.
It is then easy to calculate the first  moments of $n_{g+1}$
\begin{eqnarray}
\langle n_{g+1} \rangle = N \left[1-\left(1 - {1 \over N} \right)^{2 n_g}
\right],\\ 
\fl\langle n_{g+1}^2\rangle - \langle n_{g+1} \rangle^2 = N\left(1 - {1
\over N} \right)^{2 n_g} + N (N-1) \left(1 - {2 \over N} \right)^{2 n_g}
- N^2 \left(1 - {1 \over N} \right)^{4 n_g} .
\end{eqnarray}

For large $N$,  one can see that
\begin{equation}
\langle n_{g+1} \rangle \simeq N \left(1-   \exp \left[- {2  n_g\over N}
\right]   \right),
\label{deter}
\end{equation}
 and that 
\begin{equation}
\fl
\langle n_{g+1}^2\rangle - \langle n_{g+1} \rangle^2  \simeq
\cases{
N \exp \left[- {2  n_g\over N}  \right]  - (N+2 n_g) \exp \left[- {4
n_g\over N}  \right]   &for $n_g \sim N$,
\\ {2 n_g^2 - n_g \over N}&for $n_g \ll N$.
}
\end{equation}
These expressions show that, for large $N$, the fluctuations of $n_{g+1}$ are  small compared to the average $\langle n_{g+1} \rangle$
and the  evolution of $n_g$ is very well approximated by a deterministic evolution
\begin{equation}
n_{g+1} = N  F\left({n_g \over N}\right)\quad\mbox{ and }\quad
n_0=1,\quad\mbox{with}\quad
F(x) = 1-e^{-2 x}  .
\label{F-def}
\end{equation}
As long as $n_g/N \ll 1$, the function $F(x)$ can be approximated by
a linear function $F(x) \simeq 2x$, and one has $n_g  \simeq   2^g$ for
$g \ll \ln  N$.
On the other hand,  in the long time limit,  $ n_g \to N x^*$ where
$x^*\simeq0.8$ is the  attractive fixed point of the map $x \to F(x)$.
The meaning of $ x^*$ is simply the fraction of the population whose
lineage does not become extinct after many generations, while $1-x^*$
is the fraction of the population whose lineage becomes extinct.

The map $x \to F$ allows one to determine several  other properties of the genealogies.
For example if one tries  to compare the genealogies  of $k$ individuals
and calls $n_g(k)$ the total number of distinct ancestors of at least one
of these $k$ individuals
at generation~$g$ in the past, it is clear that up to a change of initial
condition  $n_g(k)$ evolves as $n_g$ in~(\ref{F-def}):
\begin{equation}
n_{g+1}(k)= N F\left({n_g(k) \over N}\right)\quad\mbox{ with }\quad
n_0(k)=k.
\label{ng}
\end{equation}
Again, as long as $g \ll \ln  N$, one finds $n_g(k)=k\,2^g$ and the
ancestors of the $k$ individuals are all distinct. On the other hand, all
the $n_g(k) $ converge to the same value $N x^*$ meaning that all
ancestors become common to the whole population.

The number of ancestors $m_g(2)$ common  to $2$ individuals at
generation $g$ in the past can be written as
\begin{equation}
m_g(2) = 2 n_g(1) - n_g(2) .
\nonumber
\end{equation}
More generally  the number of ancestors $m_g(k)$ common  to $k$ individuals at generation $g$ 
\begin{equation}
 m_g(k) = \sum_{p=1}^k  {k! \over p! \ (k-p)!} (-1)^{p+1} n_g(p) .
\label{mg}
\end{equation}

For $N$ large and $k \ll N$, one can expand the solution of (\ref{ng}) in powers of $1/N$  to get
\begin{equation}
\fl
n_g(k)= k\,2^g  - {k^2 \over N} \left( 2^{2 g} - 2^g \right) + 
 {k^3 \over 9 N^2} \left(  8 \times 2^{3 g}  -18 \times  2^{2 g} +10 \times
2^g \right) +  \cdots 
\label{expansion}
\end{equation}
Using this expansion in (\ref{mg}), one gets that, as long as  $k\,2^g \ll N$, 
\begin{equation}
 m_g(k) \sim {2^{k g} \over N^{k-1}} .
\nonumber
\end{equation}
By requiring that $m_g(p) \sim 1$,  one then obtains the number $g_k$ of
generations needed to find at least one common ancestor to $k$
individuals
\begin{equation}
g_k \simeq {k-1 \over k} {\ln N \over \ln 2}  .
\label{gk}
\end{equation}
This expression agrees well with simulations performed by Stephane Munier \cite{Munier}. As the size $k$ increases, it converges to Chang's expression (\ref{G1}).

\bigbreak

For $2^g \sim N$, the expansion (\ref{expansion}) can be rewritten to leading order as a scaling function
\begin{equation}
n_g(k) \simeq N H\left( {k\,2^g \over N}  \right) \ \ \  {\rm with } \ \ \
H(x)=x-x^2 + {8 \over 9} x^3 - {46 \over 63}  x^4 + \cdots
\label{ngH}
\end{equation}
where $H(x)$ is the solution of
 \begin{equation}
H(2 x) = F[H(x)]
\label{HF}
\end{equation}
which starts at $x=0$ as $H(x)=x$ and where $F$ is given in~(\ref{F-def}).
For large $x$, one can see from (\ref{HF}) that $H(x) \to x^*$, the attractive fixed point of the map  $x \to F(x)$,  and linearizing  the map (\ref{F-def})  around this fixed point one gets that for large $x$

\begin{equation}
 H(x)  \simeq x^* - B\left({\ln x \over \ln 2}\right) x^{-\alpha} \ \ \
 {\rm where} \ \ \  \alpha = -{\ln (2 (1-x^*)) \over \ln 2}\simeq 1.3,
\label{H-asympt}
\end{equation}
 and  where $B$ is a periodic function  $B(x)=B(x+1)$ of period 1.

To find the number of generations $\widetilde{g_k}$ for a group of $k$
individuals to have all their ancestors in common, one should write that
$n_g(k)-m_g(k)=0$. Here, as we use deterministic equations, the
difference $n_g(k)-m_g(k)$ tends exponentially to zero without ever
reaching it, so we replace this condition by $n_g(k)-m_g(k)\sim 1$
arguing that the stochastic nature of the evolution of $n_g(k)$ and
$m_g(k)$  should make this difference  vanish
quickly after it has become of order $1$. 
From (\ref{ngH},\ref{H-asympt}),
\begin{equation}
n_g(k)  \simeq N H\left({ k \  2^g
\over N}\right) \simeq N \left[x^* -
\frac{N^\alpha}{2^{g\alpha}}\,\frac1{k^\alpha}B\left({\ln k - \ln N \over \ln
2}\right)  \right],
\label{ngB}
\end{equation}
and one then gets (\ref{mg}) for  $m_g(k)$ 
\begin{equation}
m_g(k) \simeq N \left[x^* -
\frac{N^\alpha}{2^{g\alpha}}\sum_{p=1}^k\frac{k!}{p!(p-k)!}(-1)^{p+1}
\frac{1}{p^\alpha} B\left({\ln p - \ln N \over \ln
2}\right)  \right].
\label{mgB}
\end{equation}
Then, as  the function $B$ is periodic, the coefficient to $N^\alpha$ 
remains bounded in (\ref{ngB},\ref{mgB})
as $N$ becomes large and one gets that $n_g(k)-m_g(k)$ is $N^{\alpha+1} /
2^{g\alpha}$ times a number of order~1. Therefore, one gets for
$\widetilde{g_k}$:
\begin{equation}
\widetilde{g_k}  \simeq {(1+ \alpha) \ln N  \over  \alpha \ln 2} =
\left[{1 \over \ln 2} - {1 \over \ln (2(1-x^*))} \right] \ln N.
\label{gk-tilde}
\end{equation}
We see that  for large $N$, the times $\widetilde{g_k}  $ do not depend
on $k$ to leading order in $N$ and that the expression agrees with Chang's prediction (\ref{G2}).

\subsection{An attempt to include selection}
One can imagine various ways of introducing selection in a model with
sexual reproduction. In an attempt to do so, we consider the same model
as in the neutral case: a population of fixed size $N$,   each individual
having  its two parents chosen at random in the previous generation. Then
we introduce selection by saying that  each parent $i$ is chosen with a
probability $p_i $. We assume that these $p_i$ are all of order $1/N$,
that their distribution  remains the same at all generations, and  that  there is no
correlation between the $p_i$ of an individual and the $p_i$'s of its
parents (the selective advantage is not inheritable). 
All the analysis of  the neutral case can be extended to this case, the only change being that the function $F(x)$ defined in (\ref{F-def}) becomes
\begin{equation}
F= {1 \over N} \sum_{i=1}^N \left( 1-e^{- 2 N  p_i x}\right)  \ \ \ {\rm
with} \ \ \  \sum_i p_i= 1.
\end{equation}
As the function $F$ is modified, all the properties which depend on the  precise shape  of $F$,   such as the fixed point $x^*$, are modified.
In particular (\ref{G2},\ref{gk-tilde}) become
\begin{equation}
\widetilde{g_k} \simeq
\widetilde{G_N}  \simeq 
\left[{1 \over \ln 2} - {1 \over  \ln F'(x^*)} \right] \ln N,
\end{equation}
while the times $g_k$ to find at least one common ancestor remain unchanged (\ref{gk}).

\begin{figure}
\begin{indented}\item[]
\includegraphics[width=.75\textwidth]{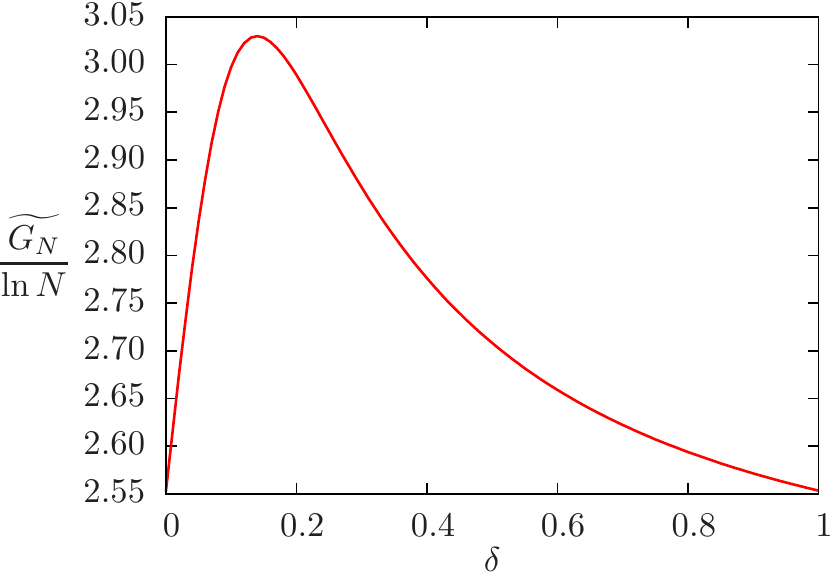}
\end{indented}
\caption{ The time $\widetilde{G_N}$ for the whole population to have all their ancestors in common versus $\delta$ when the $p_i$ take two values: ($1/(1+4 \delta)$ with probability $1-\delta$ and  $5/(1+4 \delta)$ with probability $\delta$.
We see that the effect of selection (here having two possible values  of the $p_i$ is to increase the time  $\widetilde{G_N}$ while in the limit $\delta=0$ or $\delta=1$ one recovers Chang's value (\ref{G2}).}
\label{rs}
\end{figure}

For a particular choice of the $p_i$ where $  p_i= 5 /(1+ 4 \delta)/N$
for $i< N  \delta$ and $p_i=1/(1+  4 \delta)$/N for $i > N\delta $
we show  the $\delta$ dependence of $ \widetilde{G_N} $ on figure \ref{rs}.
We see that the time for  the whole population to have all their ancetors in common  is modified by selection and that, at least  for the particular distribution considered here, selection has the effect of increasing the time  $
\widetilde{g_k} \simeq
\widetilde{G_N} $. This is somewhat surprising as it  goes in opposite direction of what we saw in the case of asexual
evolution where the effect of selection was to shorten the coalescence times.

\section{Conclusion}

Here we have reviewed a series of recent results on  simple 
models of evolution with and without selection.

In the case of asexual evolution, at least for the models we considered
here in sections 3 and 4, the effect  of selection is to make the
coalescence times $T_p$ (=the age of the most recent common ancestor to
$p$ individuals) grow logarithmically (\ref{T2-expo},\ref{ln3}) with the
size of the population in contrast to the linear growth of the (neutral)
Wright-Fisher model (section~2.2).  Moreover the statistics of the
genealogies are modified by selection and  seem to be always given by
the Bolthausen-Sznitman coalescent~(\ref{bs}).

Apart for the exponential model, for which a full mathematical treatment 
is possible (see section 3 and \cite{BDMM2.06,BDMM.07,BrunetDerrida.12}), a detailed theory, showing that 
the generic models of section~4 lead also to the Bolthausen-Sznitman 
coalescent,  is  still missing. Recent mathematical works, however, where 
selection is replaced by an absorbing wall moving at a constant velocity
\cite{Kesten.78,DerridaSimon.07,BerestyckiBerestyckiSchweinsberg.10}
or with a more complicated dynamics \cite{Maillard.11} to keep the size of the population 
bounded have shown that the Bolthausen-Sznitman coalescent does give the 
tree statistics one should see for this whole class of models.

When looking at the $N$ values $x_i(g)$ as a cloud of points moving along 
the fitness axis, all the models  in section 4 lead to a motion of this 
cloud of points very related to the motion of travelling waves of the 
noisy Fisher-KPP equation. In fact the same phenomenological theory \cite{BDMM.06,BDMM.07} which 
allows one to unsderstand the fluctuations of the position of these 
travelling waves can be used for the models of section~4 and leads to the 
Bolthausen-Sznitman statistics of the genealogies of these models.

Other models of evolution with selection on a smooth fitness 
landscape have been studied in the past
\cite{Kessler.97,DesaiFisherMurray.07,RouzineBrunetWilke.08}.
In these models each individual has an average number of offspring proportional
to the exponential of its fitness (very much like in the exponential model of section \ref{expo-mod}) giving a huge advantage to the leaders.
Obviously, an interesting open 
question would be to investigate their  genealogies  in order to  see 
whether these other models  lead to new  statistics or if  they belong to 
the Bolthausen-Sznitman universality class.
Whether other ways of introducing selection (rugged landscapes
\cite{JainKrug.05}, competition between a few alleles \cite{NeuhauserKrone.97})
could lead to a few number of other universality classes is also an interesting question to investigate.

In the sexual case, we have recalled the main results due to Chang
\cite{Chang.99}, on the 
numbers of generations needed to find at least one common ancestor or all 
common ancestors to the  whole population in the neutral case. In an 
attempt to include selection, we have seen  how these times are modified, 
with the rather counter-intuitive result that selection might increase the 
number of generations to have all ancestors common to the whole
population 
(in contrast to what happens in the asexual case where this time decreases due to selection).
Of course a more exhaustive study of the effect of selection on the 
genealogies, in the case of sexual reproduction, would be suitable
in particular when one includes recombination
\cite{BartonEtheridge.04,NeherShraimanFisher.10,HudsonKaplan.88}.

\ack
We wish to thank S. Munier and A.H. Mueller with whom several  results reviewed here were obtained.

\end{document}